\documentclass[a4paper,cite]{article}
\usepackage{latexsym}

\usepackage[usenames,dvipsnames]{pstricks}
\usepackage{pst-grad} % For gradients
\usepackage{pst-plot} % For axes

\newcommand{\lbl}[1]{\label{eq:#1}}
\newcommand{ \rf}[1]{(\ref{eq:#1})}

\newcommand{\be}{\begin{equation}}
\newcommand{\ee}{\end{equation}}
\newcommand{\bea}{\begin{eqnarray}}
\newcommand{\eea}{\end{eqnarray}}
\newcommand{\setl}{\setlength\arraycolsep{2pt}} 

\newcommand{\noi}{\noindent}

\newcommand{\ra}{\rightarrow}
\newcommand{\Ra}{\Rightarrow}

\newcommand{\cF}{{\cal F}}

\newcommand{\cM}{{\cal M}}

\newcommand{\cO}{{\cal O}}

\newcommand{\Imm}{\mbox{\rm Im}}
\newcommand{\Ree}{\mbox{\rm Re}}

\newcommand{\GeV}{\mbox{\rm GeV}}

\newcommand{\annd}{\mbox{\rm and}}
\newcommand{\foor}{\mbox{\rm for}}

\input epsf
\usepackage{latexsym}

\usepackage[T1]{fontenc}

\usepackage[latin1]{inputenc}

\usepackage{graphicx}
\usepackage{color,pst-plot}

\usepackage{amsmath}
\usepackage {amsfonts,amssymb,amstext}
\def\theequation{\arabic{section}.\arabic{equation}}

\allowdisplaybreaks[1]
\begin{document}

\begin{titlepage}

\begin{flushright}
\today  \\
\end{flushright}

\vspace*{0.2cm}
\begin{center}

{\LARGE\bf 
On Constraints Between $\Delta\alpha_{\rm had}(M_Z^2)$ and $(g_{\mu}-2)_{\rm HVP}$}\\[2cm]

{\large\bf Eduardo de Rafael}\\[0.75cm]

  {\it  Aix-Marseille Univ, Universit\'e de Toulon, CNRS, CPT, Marseille, France} 
			
\end{center}

\vspace*{2.0cm}   

\begin{abstract}
 
{ I discuss the  possibility of estimating the shift $\delta\Delta\alpha(M_Z^2)$ on the running of the EM-coupling to the $M_Z$ scale, induced by the discrepancy $\Delta a_{\mu}$  between two precise determinations of the hadronic vacuum  polarization contribution to $g_{\mu}-2$. It is shown that  the size of $\Delta a_{\mu}$  implies rigorous  bounds on the $\delta\Delta\alpha(M_Z^2)$-shift. Any extra contribution to this minimal shift necessarily depends on the specific shape of the underlying spectral function $\frac{1}{\pi}\Imm\Delta(t)$ responsible for the $\Delta a_{\mu}$-discrepancy. I show that, in the case of  a  quark model, the total $\Delta\alpha(M_Z^2)$-shift remains small. I also discuss the scenario where $\frac{1}{\pi}\Imm\Delta(t)$  is a constant in a finite $t$-region and show that in this case, up to a $t_{\rm max}\precsim 5~\GeV^2$, the size of the total $\delta\Delta\alpha_{\rm had}^{(5)}(M_Z^2)$-shift, remains below or of the order of  the present error value on $\Delta\alpha_{\rm had}^{(5)}(M_Z^2)$.}
 
\end{abstract}

\end{titlepage}

\section{\large Introduction.}
\setcounter{equation}{0}
\def\theequation{\arabic{section}.\arabic{equation}}

\noi 
Testing the EW-Standard Model requires knowledge of the running of the EM-coupling induced by the hadronic interactions, from the low-energy measured value $\alpha = 1/137.036...$  up to the scale of the  $M_Z$-mass: $M_{Z}=91.1876\pm 0.0021~\GeV$. This running is encoded in a quantity called $\Delta\alpha_{\rm had}(M_Z^2)$, which in terms of the hadronic spectral function $\frac{1}{\pi}\Imm\Pi_{\rm had}(t)$ is given by the principal value integral:
\be\lbl{eq:delta}
\Delta\alpha_{\rm had}(M_Z^2)={\rm PV}\!\! \! \int_{t_0=4m_{\pi}^2}^\infty \frac{dt}{t}\ \frac{M_Z^2}{M_Z^2 -t}\ \frac{1}{\pi}\Imm\Pi_{\rm had}(t)\,,
\ee
and contributes to the running  of $\alpha$ at the $M_Z$-scale via the relation:
\be
\alpha(M_Z^2)=\frac{\alpha}{1-\Delta\alpha_{\rm had}(M_Z^2)-\Delta\alpha_{\rm lep}(M_Z^2)}\,,
\ee
where 
\be
\Delta\alpha_{\rm lep}(M_Z^2)=(314.979\pm 0.002)\times  10^{-4}\
\ee
is the contribution due to the EM-couplings of the leptons~\cite{Stein98,Sturm13}.

The same hadronic spectral function governs the hadronic vacuum polarization (HVP) contribution to the muon anomalous magnetic moment via the equivalent integral representations~\cite{BM61,   GdeR69, LPdeR72, EdeR94}:

{\setl
\bea\lbl{eq:str}
 a_{\mu}^{\rm HVP} &  = &  \frac{\alpha}{\pi}\int_{t_0}^{\infty}
\frac{dt}{t}\underbrace{\int_{0}^{1}dx\frac{x^2(1-x)}{x^2+\frac{t}{m_{\mu}^2}(1-x)}}_{K(t)}
\frac{1}{\pi}\Imm\Pi_{\rm had}(t)\,,\\
 & = & -\frac{\alpha}{\pi}\int_0^1 dx (1-x)\ \Pi_{\rm had}\left(\frac{x^2}{1-x}m_{\mu}^2 \right)\,, \lbl{eq:euc}
\eea}

\noi
where  
\be\lbl{eq:disperr}
-\Pi_{\rm had}(Q^2)=\int_{t_0}^\infty \frac{dt}{t} \frac{Q^2}{Q^2 +t}\ \frac{1}{\pi}\Imm\Pi_{\rm had}(t)\,,\quad Q^2\equiv \frac{x^2}{1-x}m_{\mu}^2\,.
\ee
is the hadronic self-energy in the Euclidean and the optical theorem relates the spectral function $\frac{1}{\pi}\Imm\Pi_{\rm had}(t)$  to the observable one-photon annihilation  cross-section:
\be
\sigma(t)_{e^+e^- \ra {\rm had}}\underset{{m_e\ra 0}}{\thicksim}\frac{4 \pi^2 \alpha }{t} \frac{1}{\pi}\Imm\Pi_{\rm had}(t)\,.
\ee
This is the way that experimental data-driven  determinations of $\Delta\alpha_{\rm had}(M_Z^2)$ and $a_{\mu}^{\rm HVP}$
have been obtained~(see refs.~\cite{DHMZ20,KNT20} and the recent review in ref.~\cite{WB20}). 

In ref.~\cite{KNT20}, the  contribution to $\Delta\alpha_{\rm had}(M_Z^2)$ from the hadronic flavours induced by  $u,d,s,c,b$ quarks called $\Delta\alpha_{\rm had}^{(5)}(M_Z^2)$,  has been obtained~\footnote{For other determinations of $\Delta\alpha_{\rm had}(M_Z^2)$, all compatible within errors, see e.g. Table~6 in this reference and ref.~\cite{WB20}.} using  experimental data up to
\be\lbl{eq:scaleex}
\sqrt{t_{\rm max}}=11.199~\GeV\,,
\ee
and pQCD beyond that energy with the overall result
\be\lbl{eq:dexp}
\Delta\alpha_{\rm had}^{(5)}(M_Z^2)=(276.09\pm 1.12)\times 10^{-4}\,.
\ee

{ 
On the other hand, there is a  recent LQCD determination of $a_{\mu}^{\rm HVP}$ reported by the BMWc collaboration~\cite{BMWc20}, which differs significantly from the data-driven results on $\sigma(t)_{e^+e^- \ra {\rm had}}$  quoted in refs.~\cite{DHMZ20,KNT20}. To be precise, let me compare the lowest order HVP result of ref.~\cite{KNT20} (which has the largest discrepancy)
 with the one in ref.~\cite{BMWc20} i.e.,
\be\lbl{eq:g2ex}
 a_{\mu}^{\rm HVP}({\rm KNT})=(692.78\pm 2.42)\times 10^{-10}\,,\quad \rm{versus} \quad 
{ a_{\mu}^{\rm HVP}({\rm BMWc})=(708.7\pm 5.3)\times 10^{-10}}\,.
\ee
This  discrepancy
\be\lbl{eq:discr}
{ \Delta a_{\mu}\equiv a_{\mu}^{\rm HVP}({\rm BMWc})- a_{\mu}^{\rm HVP}({\rm KNT})=(15.92\pm 5.83)\times 10^{-10}}
\ee
of {  $2.7\sigma$},
 has  prompted  a reexamination  of the question raised sometime ago~\cite{PMS08}:

{\it At what precision is $\Delta\alpha_{\rm had}^{(5)}(M_Z^2)$ constrained by an accurate determination of $a_{\mu}^{ \rm HVP}$~?}

\noi
Possible scenarios have been discussed in refs.~\cite{CHMM20,KMPS20} that, though model dependent, suggest potential inconsistencies between the $a_{\mu}^{\rm HVP}({\rm BMWc})$ result and the present value for $\Delta\alpha_{\rm had}^{(5)}(M_Z^2)$ in Eq.~\rf{eq:dexp}~\footnote{{ See also ref.~\cite{MS20} which has appeared after the completion of this work.}}}.

{ The purpose of this note is to address the question above from a  new perspective. In the next section, I discuss a model-independent analysis of the observables $a_{\mu}^{\rm HVP}$ and  $\Delta\alpha_{\rm had}^{(5)}(M_Z^2)$ within the Mellin-Barnes Representation framework. Section III discusses  properties of the {\it first moment}  related to these two observables. Rigorous bounds on the size of the shift implied on $\Delta\alpha_{\rm had}^{(5)}(M_Z^2)$ by the $\Delta a_{\mu}$-discrepancy, {\it when only the value of $\Delta a_{\mu}$ is known},  are then derived in Section IV.
Alternative phenomenological { models} to those discussed in ref.~\cite{CHMM20} are then presented in Section V with conclusions in Section VI.}

\vspace*{0.5cm}

\section{\large Mellin-Barnes Representations.}
\setcounter{equation}{0}
\def\theequation{\arabic{section}.\arabic{equation}}

\noi
Mellin-Barnes representations are particularly useful to obtain asymptotic expansions of functions. In the case of $a_{\mu}^{\rm HVP}$~\cite{EdeR14}:
\be\lbl{eq:MBg}
a_{\mu}^{\rm HVP} =    \left(\frac{\alpha}{\pi}\right) \frac{m_{\mu}^2}{t_0}\frac{1}{2\pi i}\int\limits_{c_s -i\infty}^{c_s +i\infty}ds\left(\frac{m_{\mu}^2}{t_0} \right)^{-s} \cF(s)\ { \cM(s)}\,,\quad  c_s \equiv \Ree(s) \in ]0,1[\,, 
\ee
where $ \cF(s)  =  -\Gamma(3-2s)\ \Gamma(-3+s)\ \Gamma(1+s)\,,$ and
\be
\cM(s)=\int_{t_0}^\infty \frac{dt}{t}\left( \frac{t}{t_0}\right)^{s-1} \frac{1}{\pi}\Imm\Pi_{\rm had}(t)\,,
\ee
is the Mellin transform of the spectral function. 
This representation has been proposed as a possible way to obtain successive approximations to $a_{\mu}^{\rm HVP}$ from the knowledge of the moments $\cM(s)$ at $s=0,-1,-2 \cdots$ and their extrapolation at all $s$-values using Mellin-Barnes approximants (see refs.~\cite{EdeR17,ChGdeR18} and references therein).

There is also  
a similar representation for $\Delta\alpha_{\rm had}(M_Z^2)$:
\be\lbl{eq:MBD}
\Delta\alpha_{\rm had}(M_Z^2)= \frac{1}{2\pi i}\int\limits_{c_{s}-i\infty}^{c_{s}+i\infty} ds\ \int_{t_0}^\infty \frac{dt}{t}\frac{1}{\pi}\Imm\Pi_{\rm had}(t)\ \left(\frac{t}{M_Z^2} \right)^{-s}\Gamma(s)\ \Gamma(1-s)\frac{\pi}{\Gamma\left(\frac{1}{2} +s\right)\Gamma\left(\frac{1}{2}-s\right)}\,,
\ee
which differs by the factor $\frac{\pi}{\Gamma\left(\frac{1}{2} +s\right)\Gamma\left(\frac{1}{2}-s\right)}$~\cite{DG20}, when 
compared to the representation in the Euclidean:
\be
\Delta\alpha_{\rm had}(-M_Z^2)= \frac{1}{2\pi i}\int\limits_{c_{s}-i\infty}^{c_{s}+i\infty} ds\ \int_{t_0}^\infty \frac{dt}{t}\frac{1}{\pi}\Imm\Pi_{\rm had}(t)\ \left(\frac{t}{M_Z^2} \right)^{-s}\Gamma(s)\ \Gamma(1-s)\,.
\ee
Restricting the integration over the spectral function to un upper scale $t_{\rm max}$, as  we shall later do, one has
\be\lbl{eq:mbint}
\Delta\alpha_{\rm had}^{(5)}(M_Z^2)_{\rm data}= \frac{1}{2\pi i}\int\limits_{c_{s}-i\infty}^{c_{s}+i\infty} ds \left(\frac{t_{\rm max}}{M_Z^2} \right)^{-s}\ \tilde{\cM}(s)\ \frac{\Gamma(s)\ \Gamma(1-s)}{\Gamma\left(\frac{1}{2} +s\right)\Gamma\left(\frac{1}{2}-s\right)}\pi\,,
\ee
where $\tilde{\cM}(s)$ is now the restricted Mellin transform
\be
\tilde{\cM}(s)=\int_{t_0}^{t_{\rm max}} \frac{dt}{t}\ \left(\frac{t}{t_{\rm max}} \right)^{-s}\frac{1}{\pi}\Imm\Pi_{\rm had}(t)\,.
\ee
This representation governs the asymptotic expansion of $\Delta\alpha_{\rm had}^{(5)}(M_Z^2)_{\rm data}$ in terms of the ratio $\left(\frac{t_{\rm max}}{M_Z^2}  \right)$.  
The series expansion in this  ratio is fixed by the successive singularities  at the left of the fundamental  strip: $0 <{\Ree~c}_s <1$,  i.e. at $s=0,-1,-2,\cdots$ of the integrand in the r.h.s. of Eq.~\rf{eq:mbint}, and the coefficients  are proportional to the moments:
\be\lbl{eq:moms}
\int_{t_0}^{t_{\rm max}} \frac{dt}{t}\ \left(\frac{t}{t_{\rm max}} \right)^{n}\frac{1}{\pi}\Imm\Pi_{\rm had}(t)\,, \quad n=0,1,2,\cdots\,,
\ee
which  give information about the spectral function  $\frac{1}{\pi}\Imm\Pi_{\rm had}(t)$ in a different region to the one provided by the moments:
\be\lbl{eq:momsg}
\int_{t_0}^{\infty} \frac{dt}{t}\ \left(\frac{t_0}{t} \right)^{n}\frac{1}{\pi}\Imm\Pi_{\rm had}(t)\,, \quad n=1,2,\cdots\,,
\ee
which govern the $(g_{\mu}-2)_{\rm had}$ contribution~\cite{EdeR14, EdeR17, ChGdeR18}. Therefore, within this perspective,  the possibility  of relating the two observables $\Delta\alpha_{\rm had}^{(5)}(M_Z^2)$ and $(g_{\mu}-2)_{\rm had}$ {  seems {\it a priori}  a rather difficult task, unless of course one uses data or a model for the hadronic spectral function, or performs  fully dedicated  LQCD evaluations of both observables.}

Moments like those  in Eq.~\rf{eq:moms} can be related to the contour integrals (where $Q^2$ is now a complex variable)~\cite{BNP92}:
\begin{equation}\lbl{eq:FESRstandard}
	\frac{1}{2\pi i}\oint_{\vert Q^2\vert =t_{\rm max}}\frac{dQ^2}{Q^2}\left( \frac{Q^2}{t_{\rm max}}\right)^n 	
	\Pi_{\rm had}(Q^2)  = (-1)^{n+1}
	\int_{t_0}^{t_{\rm max}}\frac{dt}{t} \left(\frac{t}{t_{\rm max}} \right)^{n}\ \frac{1}{\pi}\Imm\Pi_{\rm had}(t)\,,
\end{equation}
which define a particular case  of Finite Energy Sum Rules. The experimental determination of the integrals on the r.h.s. can  thus be confronted, for sufficiently large $t_{\rm max}$-values,  to a  theoretical calculation of the l.h.s. using pQCD and the OPE at large-$Q^2$values.  

By contrast, the moments in Eq.~\rf{eq:momsg} are related to derivatives of the self-energy function $\Pi_{ \rm had}(Q^2)$ at $Q^2 =0$~\cite{EdeR14}: 
\be\lbl{eq:MasterEq}
\frac{(-1)^{n+1} }{(n+1)!}( t_0)^{n+1} \underbrace{\left(\frac{\partial^{n+1}}{(\partial Q^2 )^{n+1}}\Pi(Q^2)\right)_{Q^2 =0}}_{\rm  LQCD}=
\underbrace{\int\limits_{t_0}^{\infty}\frac{dt}{t}\left(\frac{t_0}{t} \right)^{1+n}
\frac{1}{\pi}\Imm\Pi(t)}_{\rm Experiment}\,,\quad n=0,1,2,\cdots\,.
\ee
As discussed  in refs.~\cite{ EdeR14,EdeR17,ChGdeR18} they provide excellent tests to confront LQCD evaluations of the l.h.s. with experimental results, but they invoke a totally different $Q^2$-region to the one which applies to Eq.~\rf{eq:FESRstandard}.

\vspace*{0.5cm}

\section{\large  The First Moment.}
\setcounter{equation}{0}
\def\theequation{\arabic{section}.\arabic{equation}}

\noi

I want to focus attention on    
the first moment $n=0$ in Eq.~\rf{eq:MasterEq}:
\be\lbl{eq:momentss}
\left. -t_{0}\frac{\partial \Pi_{\rm had}(Q^2)}{\partial Q^2}\right\vert_{Q^2 =0}=\int\limits_{t_0}^{\infty}\frac{dt}{t}\ \frac{t_{0}}{t}\ 
\frac{1}{\pi}\Imm\Pi_{\rm had}(t)\,.
\ee
A long time ago, the authors of ref.~\cite{BdeR69} showed that there is an upper and lower bound for $a_{\mu}^{\rm HVP}$ in terms of this moment, namely:
\be\lbl{eq:BdeR}
\frac{\alpha}{\pi}\frac{1}{3}\frac{m_{\mu}^2 }{t_{0}}\left(\ \int\limits_{t_0}^{\infty}\frac{dt}{t}\frac{t_{0}}{t}
\frac{1}{\pi}\Imm\Pi_{\rm had}(t)\right) \left[1-  f\left(\frac{t_0}{m_{\mu}^2}\right) \right]< a_{\mu}^{\rm HVP} <\frac{\alpha}{\pi}\frac{1}{3}\frac{m_{\mu}^2}{t_{0}}\left(\  \int\limits_{t_0}^{\infty}\frac{dt}{t}\frac{t_{0}}{t}
\frac{1}{\pi}\Imm\Pi_{\rm had}(t)\right)
\,,
\ee
where 
\be\lbl{eq:BdeRbound}
f\left(\frac{t_0}{m_{\mu}^2}\right)=3\ \int_0^1 dx\frac{x^4}{x^2+\frac{t_0}{m_{\mu}^2}(1-x)}=0.36655\,,\quad \foor\quad t_0=4m_{\pi^{\pm}}^2\,.
\ee
The bounds follow from the observation  that the kernel $K(t)$ in Eq.~\rf{eq:str} can be written as follows:
\be
K(t)=\frac{m_{\mu}^2}{t}\int_0^1 dx \ [ x^2  -\underbrace{\left(\frac{x^4}{x^2 +\frac{t}{m_{\mu}^2}(1-x)}\right)}_{\frac{1}{3}f \left(\frac{t}{m_{\mu}^2}\right)}]\,,
\ee
and the fact that, within the $t$-integration range $t_0 < t < \infty$,
\be
0 < f\left(\frac{t}{m_{\mu}^2}\right)<f\left(\frac{t_{0}}{m_{\mu}^2}\right)\,.
\ee
In the language of the Mellin-Barnes representation of Section II, the upper bound in Eq.~\rf{eq:BdeR} corresponds to retaining only the leading term in the singular expansion of $\cF(s)$, i.e. the contribution from the simple pole at $s=0$. This contribution can be seen as induced by an effective local operator $\partial^{\lambda}F^{\mu\nu}\partial_{\lambda}F_{\mu\nu}$ to which I shall later come back. The other terms of the singular expansion of $\cF(s)$ give successive corrections to the upper-bound. The simple poles generate the moments in Eq.~\rf{eq:MasterEq} for $n\ge 1$, which can also be seen as induced by local operators of higher and higher dimension. The double poles  generate the  log-weighted moments discussed in ref.~\cite{EdeR14}:
\be\lbl{eq:momslogs}
\int_{t_0}^{\infty} \frac{dt}{t}\ \left(\frac{t_0}{t} \right)^{n}\ \log \frac{t}{t_0}\ \frac{1}{\pi}\Imm\Pi_{\rm had}(t)\,, \quad n=1,2,\cdots\,,
\ee
and can be seen as generated by successive non-local operators.
The lower bound in Eq.~\rf{eq:BdeR} is a rigorous  bound to the contribution  from the sum of  all the terms of the singular expansion of $\cF(s)$ beyond the leading one and, therefore, to the total effect of the underlying string  of local and non-local operators which give rise to these terms. The remarkable feature about the simplicity of the resulting lower bound in Eq~\rf{eq:BdeR} is that {\it it is  proportional to  the same leading moment which fixes the upper bound~!} Therefore, this lower bound can also be seen as induced by an effective local operator of the type $\partial^{\lambda}F^{\mu\nu}\partial_{\lambda}F_{\mu\nu}$  modulated, however, by a different coupling.

In the case of the electron $(g_e -2)_{\rm HVP}$ the bounds in Eq~\rf{eq:BdeR} are impressive.
Using  the experimental determination of the lowest HVP-moment~\footnote{Private communication from the authors of ref.~\cite{KNT20}.}: 
\be\lbl{eq:1EM}
\int_{t_0}^{\infty}\frac{dt}{t}\ \frac{t_{0}}{t}\ 
\frac{1}{\pi}\Imm\Pi_{\rm had}(t)=(0.7176\pm 0.0026)\times 10^{-3}\,,
\ee
one finds:
\be
1.8550\times 10^{-12}< a_{e}^{\rm HVP}< 1.8687\times 10^{-12}\,,
\ee
which corresponds to an accuracy of 0.37\%. I recall that the lowest order HVP experimental determination in the case of the electron anomaly gives~\cite{KNT20}:
\be
a_e ^{\rm HVP}=1.8608(66)\times 10^{-12}\,.
\ee

The contribution from the HVP-moment in Eq.~\rf{eq:1EM}  to $\Delta\alpha_{\rm had}^{(5)}(M_Z^2)$ in Eq.~\rf{eq:delta} is a small fraction of the total:
\be\lbl{eq:delan}
\Delta\alpha_{\rm had}^{(5)}(M_Z^2)=
\underbrace{\int_{t_0}^\infty \frac{dt}{t}\frac{t_0}{t}\frac{1}{\pi}\Imm\Pi_{\rm had}(t)}_{7.176\times 10^{-4}}+{\rm PV}\!\! \!\int_{t_0}^\infty \frac{dt}{t} \left( \frac{M_Z^2}{M_Z^2 -t}-\frac{t_0}{t}\right)\frac{1}{\pi}\Imm\Pi_{\rm had}(t)\,,
\ee 
but, as we shall see next, it gives the key { to relate  the two observables} $\Delta\alpha_{\rm had}^{(5)}(M_Z^2)$ and $(g_{\mu}-2)_{\rm HVP}$.

\vspace*{0.5cm}

\section{\large { The Effective Operator $\partial^{\lambda}F^{\mu\nu}\partial_{\lambda}F_{\mu\nu}$ and Rigorous Bounds.}}
\setcounter{equation}{0}
\def\theequation{\arabic{section}.\arabic{equation}}

\noi
The effective operator which  governs the size of the slope at the origin of the photon self-energy, and therefore the first moment of the underlying spectral function, is the dimension ${d=6}$ operator 
\be\lbl{eq:d6}
\frac{1}{\Lambda^2}\partial^{\lambda}F^{\mu\nu}\partial_{\lambda}F_{\mu\nu}\,.
\ee
For the HVP-contribution in particular we have
\be\lbl{eq:1mom}
\left. -\frac{\partial \Pi_{\rm had}(Q^2)}{\partial Q^2}\right\vert_{Q^2 =0}=\frac{\alpha}{\pi}\frac{1}{\Lambda_{\rm had}^2}\,,
\ee
and from Eq.~\rf{eq:MasterEq}
\be\lbl{eq:hadmom}
\frac{\alpha}{\pi}\frac{t_0}{\Lambda_{\rm had}^2}=
\int_{t_0}^{\infty}\frac{dt}{t}\frac{t_0}{t}
\frac{1}{\pi}\Imm\Pi_{\rm had}(t)\,.
\ee
The experimental value of the scale $\Lambda_{\rm had}$ which follows from Eq.~\rf{eq:1EM} is
\be
\Lambda_{\rm had}=(0.502\pm 0.001)~\GeV\,,
\ee
which, as expected, is a scale of the order of the $\rho$-mass.

The $\Delta a_{\mu}$ discrepancy in Eq.~\rf{eq:discr} can be interpreted as an excess, in some (so far) unknown $t$-range, of the underlying spectral function which results in the BMWc-evaluation,   as compared to the spectral function which has been used in the KNT-evaluation i.e.,
\be\lbl{eq:KD}
\Delta a_{\mu}=\frac{\alpha}{\pi}\int_{t_{0}}^\infty\frac{dt}{t}\ K(t)\ \frac{1}{\pi}\Imm\Delta(t)\,,
\ee
where
\be\lbl{eq:Delta}
\frac{1}{\pi}\Imm\Delta(t)\equiv \frac{1}{\pi}\Imm\Pi_{\rm BMWc}(t)-\frac{1}{\pi}\Imm\Pi_{\rm KNT}(t)\ge 0\quad\foor\quad t_0\le t\le\infty\,.
\ee
This difference of spectral functions can be seen as generated, { at least in part}, by an effective local operator of the type in Eq.~\rf{eq:d6} which induces an effective   moment
\be\lbl{eq:momd}
\frac{\alpha}{\pi}\frac{t_0}{\Lambda_{\rm eff}^2}=
\int\limits_{t_0}^{\infty}\frac{dt}{t}\frac{t_0}{t}
\frac{1}{\pi}\Imm\Delta(t)\,,
\ee
and in terms of which, the bounds in Eq.~\rf{eq:BdeR} when applied to $\Delta a_{\mu}$, give
\be\lbl{eq:BdeRII}
\frac{\alpha}{\pi}\frac{1}{3}\frac{m_{\mu}^2 }{t_{0}}\left[1-  f\left(\frac{t_0}{m_{\mu}^2}\right) \right]\left(\frac{\alpha}{\pi}\frac{t_0}{\Lambda_{\rm eff}^2}\right) < \Delta a_{\mu} <\frac{\alpha}{\pi}\frac{1}{3}\frac{m_{\mu}^2}{t_{0}}\left(\frac{\alpha}{\pi}\frac{t_0}{\Lambda_{\rm eff}^2}\right)
\,.
\ee
Including the error of $\Delta a_{\mu}$ in Eq.~\rf{eq:discr} so as to take the largest and smallest values of the discrepancy, fixes the extreme values of the scale $\Lambda_{\rm eff}^2$ to be within the limits:
\be\lbl{eq:numbounds}
{ \Lambda_{\rm eff}^2=19.89~\GeV^2~ ({\rm from~the~upper~bound})\quad\annd\quad \Lambda_{\rm eff}^2=5.85~\GeV^2 ({\rm from~the~lower~bound})} \,.
\ee
These two scales fix the range of couplings of the effective operator which is at the origin, { at least in part}, of the  anomaly discrepancy. 
Their effect on $\Delta\alpha_{\rm had}^{(5)}(M_Z^2)$ is  to add a shift to the underlined term in the r.h.s. of Eq.~\rf{eq:delan}: 

{\setl
\bea
\Delta\alpha_{\rm had}^{(5)}(M_Z^2) &  \Ra &  \underbrace{
\int_{t_0}^\infty \frac{dt}{t}\frac{t_0}{t}\frac{1}{\pi}\Imm\Pi_{\rm had}(t)+\frac{\alpha}{\pi}\frac{t_0}{\Lambda_{\rm eff}^2}}_{}+{\rm PV}\!\! \!\int_{t_0}^\infty \frac{dt}{t} \left( \frac{M_Z^2}{M_Z^2 -t}-\frac{t_0}{t}\right)\frac{1}{\pi}\Imm\Pi_{\rm had}(t) \\
 & = & \underbrace{\frac{\alpha}{\pi}\frac{t_0}{\Lambda_{\rm eff}^2}}_{\rm shift}  +{\rm PV}\!\! \!\int_{t_0}^\infty \frac{dt}{t}  \frac{M_Z^2}{M_Z^2 -t}\frac{1}{\pi}\Imm\Pi_{\rm had}(t)\,,
\eea}

\noi 
with the shift constrained by the upper and lower bounds in Eq.~\rf{eq:numbounds} to be within the limits:
\be\lbl{eq:result}
{ \underbrace{\frac{\alpha}{\pi}\frac{t_0}{19.89~\GeV^2}}_{0.091\times 10^{-4}} \le\frac{\alpha}{\pi}\frac{t_0}{\Lambda_{\rm eff}^2} \le \underbrace{ \frac{\alpha}{\pi}\frac{t_0}{5.85~\GeV^2}}_{ 0.309\times 10^{-4}}} \,.
\ee
We find, therefore, that the accepted range of the { minimal}  shift  on $\Delta\alpha_{\rm had}^{(5)}(M_Z^2)$ induced by the effective moment in Eq.~\rf{eq:momd} is rather small.  In fact,  the higher value is about { four} times  smaller than the error $1.12 \times 10^{-4}$ in the  result in Eq.~\rf{eq:dexp}. In the absence of any information about the shape of the spectral function $\frac{1}{\pi}\Imm\Delta(t)$,  this is as much as may be rigorously said about the size of the shift implied on $\Delta\alpha_{\rm had}^{(5)}(M_Z^2)$. 
 
Notice that as the discrepancy $\Delta a_{\mu}$ becomes smaller and smaller, Eq.(4.8) shows that $\Lambda_{\rm eff}^2$ becomes larger and larger and, hence, the shift in Eq.(4.11) (the first term) smaller and smaller. By contrast, as the discrepancy $\Delta a_{\mu}$ becomes larger and larger  $\Lambda_{\rm eff}^2$ becomes smaller and smaller and, hence, the shift in Eq.(4.11) larger and larger.

Any extra contribution to the minimal shift in Eq.~\rf{eq:result} is encoded in the difference
\be
{\rm PV}\!\! \!\int_{t_0}^\infty \frac{dt}{t}  \frac{M_Z^2}{M_Z^2 -t}\frac{1}{\pi}\Imm\Delta(t)-\frac{\alpha}{\pi}\frac{t_0}{\Lambda_{\rm eff}^2}\,,
\ee
with $\Imm\Delta(t)$ defined in Eq.~\rf{eq:Delta}, and its size  depends on the specific shape of the underlying BMWc-spectral function.

\vspace*{0.5cm}

\section{\large { Phenomenological { Models}.}}
\setcounter{equation}{0}
\def\theequation{\arabic{section}.\arabic{equation}}

\noi

{ \subsection{\normalsize {The Constituent Chiral Quark Model}}}

{ Models of $\frac{1}{\pi}\Imm\Delta(t)$ which lead to much larger shifts to $\Delta\alpha_{\rm had}^{(5)}(M_Z^2)$ than  { the minimal bounds} in Eq.~\rf{eq:result} have been disused in refs.~~\cite{CHMM20,KMPS20}.} 
 However, it is also possible to construct models that don't give such large shifts. The { simple} constituent chiral quark model (C$\chi$QM)~\footnote{ see ref.~\cite{GdeR12} for details}  may be used to illustrate this.  
%%%%%%%%%%%%%%%%%%%%%%%%%%%%%
\begin{figure}[!ht]
\begin{center}
%\hspace*{-1.5cm}
\includegraphics[width=0.65\textwidth]{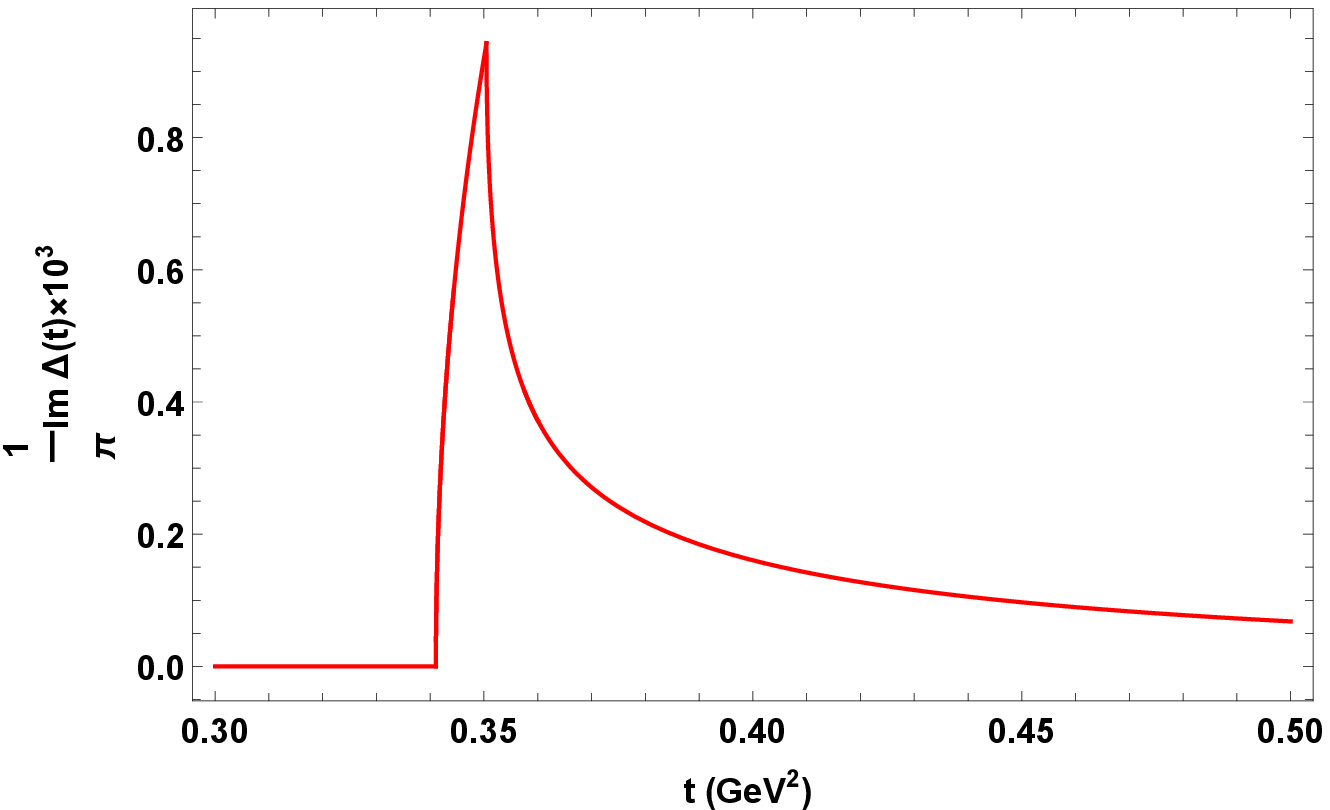} 
\bf\caption{\lbl{fig:DELTA}}
\vspace*{0.25cm}
{\it Shape of the difference of spectral functions functions in Eq.~\rf{eq:diferencespect} plotted in $10^{-3}$ units.}
\end{center}
\end{figure}
%%%%%%%%%%%%%%%%%%%%%%%%%%%%
\noi
{ The hadronic spectral function in this model:}
\be
\frac{1}{\pi}\Imm\Pi_{\chi{\rm QM}}(t,M)=\frac{\alpha}{\pi}\frac{5}{3}Nc \frac{1}{3}\left(1+\frac{2 M^2}{t}\right)
\sqrt{1-\frac{4 M^2}{t}}\theta(t-4 M^2)\,,
\ee
only depends on a $M$-mass parameter which can be adjusted so as to reproduce {  the central values of either the $a_{\mu}^{\rm HVP}({\rm BMWc})$ or  the $a_{\mu}^{\rm HVP}({\rm KNT})$ determinations}. These choices result in the values:
\be
M_{\rm BMWc}=0.292~\GeV \quad\annd\quad M_{\rm KNT}=0.296~\GeV\,,
\ee
and the relevant spectral function is then
\be\lbl{eq:diferencespect}
\frac{1}{\pi}\Imm\Delta_{\rm model}(t)= \frac{1}{\pi}\Imm\Pi_{{\chi{\rm QM}}}(t,M_{\rm BMWc})-\frac{1}{\pi}\Imm\Pi_{{\chi{\rm QM}}}(t,M_{\rm KNT})\,,
\ee
which has a shape as shown (in $10^{-3}$-units) in Fig.~1 (the cusp in the figure corresponds to the threshold opening at $4 M_{\rm KNT}^2 =0.350~\GeV^2$).
The shift  induced  on $\Delta\alpha_{\rm had}^{(5)}(M_Z^2)$ is then given by the integral
\be
{\rm PV}\!\! \! \int_{t_0=4m_{\pi}^2}^\infty \frac{dt}{t}\ \frac{M_Z^2}{M_Z^2 -t}\ \frac{1}{\pi}\Imm\Delta_{{ \rm model}}(t)=0.42\times 10^{-4}\,,
\ee
which, { as expected},  results in a number somewhat larger than the  upper bound in Eq.~\rf{eq:result}.

\noi

{ \subsection{\normalsize {Beyond the Minimal Bounds}}}

It would be nice to have some guidance {  to go beyond the minimal rigorous bounds previously discussed. In that respect I propose the following approach}:
it seems reasonable to assume that the region where the underlying spectral functions are  responsible for the $\Delta a_{\mu}$ discrepancy is a finite region in the  $t$-integration range i.e.,
\be\lbl{eq:KDres}
\Delta a_{\mu}\equiv\frac{\alpha}{\pi}\int_{t_{\rm min}}^{t_{\rm max}}\frac{dt}{t}\ K(t)\ \frac{1}{\pi}\Imm\Delta(t)\,,\quad {\rm with}\quad t_{\rm min}\ge t_0\,;
\ee
and  $t_{\rm max}<< M_Z^2$ because in any case both evaluations of the anomaly are using pQCD beyond a certain energy.
The total shift on $\Delta\alpha_{\rm had}^{(5)}(M_Z^2)$ induced by the same spectral function is then given by the integral
\be
\delta\Delta\alpha_{\rm had}^{(5)}(M_Z^2)=\int_{t_{\rm min}}^{t_{\rm max}}\frac{dt}{t}\frac{M_Z^2}{M_Z^2-t}\ \frac{1}{\pi}\Imm\Delta(t)\,,
\ee
which  is no longer a principal value integral. There is a Mellin-Barnes representation for this shift, similar to the one given in Eq.~\rf{eq:mbint}:
\be\lbl{eq:mbintres}
\delta\Delta\alpha_{\rm had}^{(5)}(M_Z^2)= \frac{1}{2\pi i}\int\limits_{c_{s}-i\infty}^{c_{s}+i\infty} ds \left(\frac{t_{\rm max}}{M_Z^2} \right)^{-s}\ \tilde{\cM}(s)\ \frac{\Gamma(s)\ \Gamma(1-s)}{\Gamma\left(\frac{1}{2} +s\right)\Gamma\left(\frac{1}{2}-s\right)}\pi\,,
\ee
where $\tilde{\cM}(s)$ is now the restricted Mellin transform of $\frac{1}{\pi}\Imm\Delta(t)$,
\be
\tilde{\cM}(s)=\int_{t_{\rm min}}^{t_{\rm max}} \frac{dt}{t}\ \left(\frac{t}{t_{\rm max}} \right)^{-s}\frac{1}{\pi}\Imm\Delta(t)\,.
\ee
{ The leading contribution in the expansion which follows from Eq.~\rf{eq:mbintres} is precisely  the one induced by the singularity of the integrand at $s=0$  and, therefore,}
\be\lbl{eq:constant}
\delta\Delta\alpha_{\rm had}^{(5)}(M_Z^2)=\tilde{\cM}(0)+\cO\left(\frac{t_{\rm max}}{M_Z^2} \right)\,,
\ee
{ with $\tilde{\cM}(0)$}  dominating largely the full contribution.

Let me { now} assume that $\frac{1}{\pi}\Imm\Delta(t)$ is a constant in the region $t_{\rm min}\le t\le t_{\rm max}$:
\be
\frac{1}{\pi}\Imm\Delta(t)\equiv \beta\ \theta(t-t_{\rm min})\ \theta(t_{\rm max}-t)\,,
\ee
something which could { well} be the case if e.g. the $\Delta a_{\mu}$ discrepancy was due to a shift in a $t$-region of the spectrum that corresponds to an offset in its height. { In this case
\be
\Delta a_{\mu}=\beta\ \frac{\alpha}{\pi}\int_{t_{\rm min}}^{t_{\rm max}}\frac{dt}{t}\ K(t)\quad\annd\quad \tilde{\cM}(0)=\beta\ \log\frac{t_{\rm max}}{t_{\rm min}}\,,
\ee
which neglecting $\cO\left(\frac{t_{\rm max}}{M_Z^2} \right)$ terms in Eq.~\rf{eq:constant}, results in a simple expression for the  total $\delta\Delta\alpha_{\rm had}^{(5)}(M_Z^2)$-shift}:
\be
\delta\Delta\alpha_{\rm had}^{(5)}(M_Z^2)=\frac{\Delta a_{\mu}}{\frac{\alpha}{\pi}\int_{t_{\rm min}}^{t_{\rm max}}\frac{dt}{t}\ K(t)}\log\frac{t_{\rm max}}{t_{\rm min}}\,.
\ee
As a  numerical  illustration of this result, I show  in Fig.~2 { the size of the $\delta\Delta\alpha_{\rm had}^{(5)}(M_Z^2)$-shift} as a function of $t_{\rm tmax}$ in the range $9m_{\pi}^2 \le t_{\rm tmax}\le 16~\GeV^2$ with $t_{\rm min}=4m_{\pi}^2$.
{  
The thick red curve in the figure shows the $\delta\Delta\alpha_{\rm had}^{(5)}(M_Z^2)$-shift corresponding to the central value of the discrepancy $\Delta a_{\mu}$ in Eq.~\rf{eq:discr}; the dashed curves the results of the same shift when including the { upper and lower errors} on $\Delta a_{\mu}$. The thick green line shows the present experimental error of $\pm 1.12$ on $\Delta\alpha_{\rm had}^{(5)}(M_Z^2)$ in Eq.~\rf{eq:dexp} in $10^{-4}$ units.} { The plot in this figure shows that up to a $t_{\rm max}\precsim 5~\GeV^2$, the size of the $\delta\Delta\alpha_{\rm had}^{(5)}(M_Z^2)$-shift, remains below or of the order of  the present error value on $\Delta\alpha_{\rm had}^{(5)}(M_Z^2)$ (the thick green line)}.

%%%%%%%%%%%%%%%%%%%%%%%%%%%%%
\begin{figure}[!ht]
\begin{center}
%\hspace*{-1.5cm}
\includegraphics[width=0.70\textwidth]{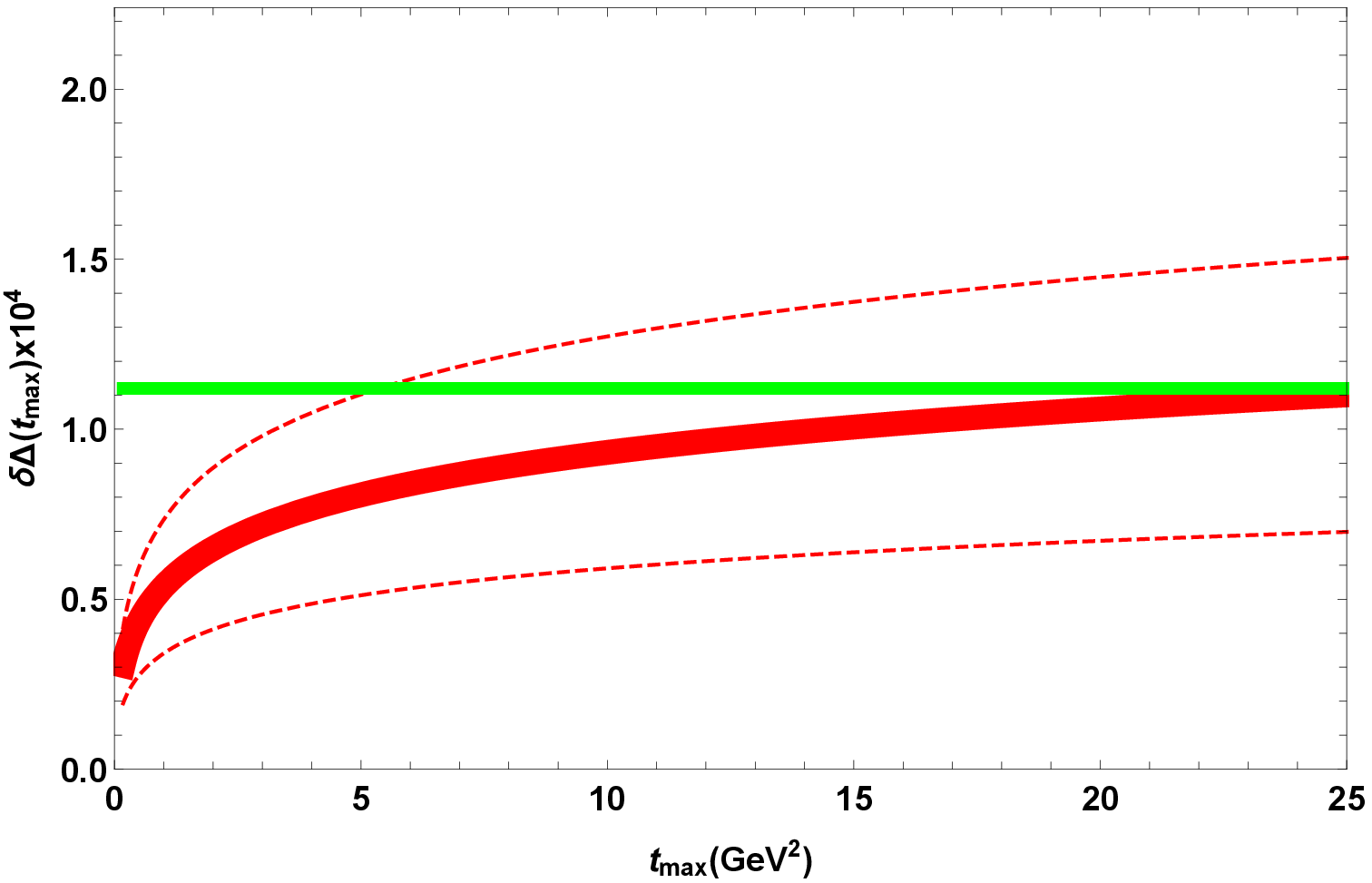} 
\bf\caption{\lbl{fig:PLOT}}
\end{center}
\vspace*{0.25cm}
{\it  Size of the $\delta\Delta\alpha_{\rm had}^{(5)}(M_Z^2)$-shift in $10^{-4}$ units as a function of the $t_{\rm max}$-choice.   
The thick red curve shows the shift corresponding to the central value of the discrepancy $\Delta a_{\mu}$; the dashed curves the  same shift when including the upper and lower errors  on $\Delta a_{\mu}$. The thick green line shows the present experimental error  on $\Delta\alpha_{\rm had}^{(5)}(M_Z^2)$, also in $10^{-4}$ units. }
\end{figure}
%%%%%%%%%%%%%%%%%%%%%%%%%%%%

\vspace*{0.5cm}

\section{\normalsize { Summary and Conclusions.}}
\setcounter{equation}{0}
\def\theequation{\arabic{section}.\arabic{equation}}

\begin{itemize}
	\item 

The bounds  in Eq.~\rf{eq:result} { are rigorous minimal bounds on the $\delta\Delta\alpha_{\rm had}^{(5)}(M_Z^2)$-shift implied by the $\Delta a_{\mu}$-discrepancy in Eq.~\rf{eq:discr}}. They do not assume any specific shape of energy dependence in the underlying spectral function $\frac{1}{\pi}\Imm\Delta(t)$  which is at the origin of the { discrepancy}. In the absence of any information about the shape of $\frac{1}{\pi}\Imm\Delta(t)$  this is as much as can be said { at present}.  

\item
{ The simple Quark Model discussed  above illustrates the fact that it is possible to construct  spectral functions which reproduce the two anomaly values $a_{\mu}({\rm BMWc})$ and $a_{\mu}({\rm KNT})$ and yet, their difference,  induces a shift on $\Delta\alpha_{\rm had}^{(5)}(M_Z^2)$ which although larger than the upper minimal bound in Eq.~\rf{eq:result}, is still much smaller than the present error in Eq.~\rf{eq:dexp}.}

\item

{ Finally, I have discussed the scenario where the underlying spectral function $\frac{1}{\pi}\Imm\Delta(t)$ responsible for the $\Delta a_{\mu}$-discrepancy  could be constant in a finite $t$-range: from threshold $t_{\rm min}=4m_{\pi}^2$ up to an arbitrary $t_{\rm max}<<M_Z^2$ and show that in this case   
the { $\Delta a_{\mu}$-discrepancy} induces  a $\delta\Delta\alpha_{\rm had}^{(5)}(M_Z^2)$-shift
which, as shown in Fig.~\rf{fig:PLOT},  remains compatible  with the present  data-driven determination of $\Delta\alpha_{\rm had}^{(5)}(M_Z^2)$ up to $t_{\rm max}\precsim 5~\GeV^2$-values.}

\end{itemize}

\vspace*{0.5cm}

{\bf Acknowledgements}

\vspace*{0.25cm}

I am very much indebted to my colleagues: Laurent Lellouch, Marc Knecht, David Greynat and J\'{e}r\^{o}me Charles, for their generous help and their   suggestions to improve the contents of this note. Discussions with the authors of ref.~\cite{CHMM20} are also acknowledged. { I also wish to thank the referee { for suggesting  improvements and clarifications}.}

\vspace*{0.5cm}
%%%%%%%%%%%%%%%%%%%%%%%%%%%%%%%%%%%%%%%%

\vfill

%\begin{flushleft 
%\fbox{Copy to {\it Marc Knecht} on 12/05/98.}
%\end{flushleft}
 
\end{document}